\documentclass[reprint,superscriptaddress,amsmath,amssymb,aps]{revtex4-1}

\usepackage{amsmath}
\usepackage{graphicx}
\usepackage{mathtools}
\usepackage{xcolor}

\def\k{{\bf k}}

\begin{document}
\title{Robust nodal behavior in the thermal conductivity of superconducting UTe$_2$}

\author{Ian M. Hayes}
\affiliation{Maryland Quantum Materials Center, Department of Physics, University of Maryland, College Park, MD 20742, USA.}

\author{Tristin E. Metz}
\affiliation{Maryland Quantum Materials Center, Department of Physics, University of Maryland, College Park, MD 20742, USA.}

\author{Corey E. Frank}
\affiliation{Maryland Quantum Materials Center, Department of Physics, University of Maryland, College Park, MD 20742, USA.}
\affiliation{NIST Center for Neutron Research, National Institute of Standards and Technology, Gaithersburg, MD 20899, USA.}

\author{Shanta R. Saha}
\affiliation{Maryland Quantum Materials Center, Department of Physics, University of Maryland, College Park, MD 20742, USA.}
\affiliation{NIST Center for Neutron Research, National Institute of Standards and Technology, Gaithersburg, MD 20899, USA.}

\author{Nicholas P. Butch}
\affiliation{Maryland Quantum Materials Center, Department of Physics, University of Maryland, College Park, MD 20742, USA.}
\affiliation{NIST Center for Neutron Research, National Institute of Standards and Technology, Gaithersburg, MD 20899, USA.}

\author{Vivek Mishra}
\author{P.J. Hirschfeld}
\affiliation{Department of Physics, University of Florida, Gainesville, FL, 32611-8440}

\author{Johnpierre Paglione}
\email{paglione@umd.edu}
\affiliation{Maryland Quantum Materials Center, Department of Physics, University of Maryland, College Park, MD 20742, USA.}
\affiliation{The Canadian Institute for Advanced Research, Toronto, Ontario, Canada.}

\begin{abstract}
      The superconducting state of the heavy-fermion metal UTe$_2$ has attracted considerable interest because of evidence for spin-triplet Cooper pairing and non-trivial topology. Progress on these questions requires identifying the presence or absence of nodes in the superconducting gap function and their dimension. In this article we report a comprehensive study of the influence of disorder on the thermal transport in the superconducting state of UTe$_2$. Through detailed measurements of the magnetic field dependence of the thermal conductivity in the zero-temperature limit, we obtain clear evidence for the presence of point nodes in the superconducting gap for all samples with transition temperatures ranging from 1.6~K to 2.1~K obtained by different synthesis methods, including a refined self-flux method.
      This robustness implies the presence of symmetry-imposed nodes throughout the range studied, further confirmed via disorder-dependent calculations of the thermal transport in a model with a single pair of nodes. In addition to capturing the temperature dependence of the thermal conductivity up to $T_c$, this model allows us to limit the possible locations of the nodes, suggesting a B$_{1u}$ or B$_{2u}$ symmetry for the superconducting order parameter. 
      Additionally, comparing the new, ultra-high conductivity samples to older samples reveals a crossover between a low-field and a high field regime at a single value of the magnetic field in all samples. In the high field regime, the thermal conductivity at different disorder levels differ from each other by a simple offset, suggesting that some simple principle determines the physics of the mixed state, a fact which may illuminate trends observed in other clean nodal superconductors.

\end{abstract}

\maketitle

Ever since the discovery of equal-spin pairing in superfluid $^3$He-A, there has been a search for an analogous equal-spin paired state in a superconductor\cite{Leggett1975}. This search was given new impetus by the realization that a triplet superconductor could host majorana zero modes that would enable fault-tolerant quantum computation.\cite{Nayak2008} 
The discovery of superconductivity in uranium ditelluride (UTe$_2$) was a watershed moment for the field because the superconducting state immediately showed two clear indications of triplet pairing. The first was that the upper critical field $H_{c2}$ exceeds the Pauli limit for all orientations. The second was that the nuclear magnetic resonance (NMR) Knight shift--which is a measure of the local spin susceptibility in the superconducting state--shows almost no change through the superconducting transition temperature $T_c$.\cite{ran_nearly_2019, Fujibayashi2022} Both of these phenomena are hard to obtain unless the electron spins are aligned in the superconducting state. 

These observations triggered extensive efforts to characterize the orbital part of the order parameter. Early measurements of the thermal conductivity and penetration depth pointed to a superconducting gap function with point nodes.\cite{metz_point-node_2019} Comparison of the penetration depth in different directions favored a multi-component gap structure with nodes away from the high symmetry axes, an observation that made sense in the context of magneto-optic Kerr rotation measurements that indicated that the superconducting order parameter broke time reversal symmetry.\cite{Ishihara2023, Hayes2021} 

In parallel, progress on the synthesis of UTe$_2$ yielded samples with notably higher $T_c$ values (up to 2~K compared to 1.6~K) and significantly reduced residual resistivity ($\rho(0K)$) values, indicating significantly reduced levels of disorder.\cite{Rosa2022, Sakai2022} Although the $H_{c2}$ values of these samples also exceed the Pauli limit, they differ in several other respects from the first generation samples. In particular, the evidence for time reversal symmetry-breaking did not reproduce \cite{Ajeesh2023} and the NMR Knight shift change through $T_c$ was observed to occur in all crystallographic directions
\cite{Matsumura2023}. These observations have been interpreted to favor a single-component order parameter, and in some cases a nodeless gap function \cite{Matsumura2023,suetsugu2023}, which in turn suggests that key features of the superconducting state were inaccessible in the older generation of samples.

Here we use measurements of the thermal conductivity $\kappa$ to investigate this possible disorder-dependent evolution of the superconducting state in UTe$_2$. We study both the finite and zero-temperature limits in samples that vary by a factor of $\sim$20 in residual resistivity, and $\sim$30\% in $T_c$. 
Throughout this range of sample quality we are able to confirm the presence of point nodes. This conclusion rests on two qualitative features in the data: the absence of a nonzero intercept in $\kappa/T$ at zero temperature, which rules out line nodes, and an immediate increase in this intercept with the application of magnetic field, which rules out a full gap. 
Our wide range of disorder levels allows us to confirm this nodal structure by performing calculations of $\kappa/T$ in a simple model of an axial $p$-wave state as functions of both temperature and magnetic field, assuming scattering of disorder and inelastic processes. The calculations are successful at capturing the evolution of the peak below $T_c$ with increasing disorder. Because thermal conductivity is direction-sensitive, these calculations also exhibit different features when the nodes are parallel to or perpendicular to the direction of current flow. Comparing these two scenarios with the data supports the conclusion that the nodes are not near the crystallographic $a$-axis.
Given our best current understanding of the Fermi surface and the current evidence in favor of a single-component order parameter, this implies a gap structure belonging to either the B$_{2u}$ or B$_{3u}$ irreducible representation of the crystal point group, consistent with conclusions reached in recent ultrasound and heat capacity experiments.\cite{theuss2023, lee2023} 
Finally, the field dependence of thermal transport in the zero temperature limit, $\kappa(H,T$$\rightarrow$0)/$T$, exhibits a crossover near 1~T to a regime where it grows at a uniform rate independent of the level of disorder, suggesting a universal behavior in the mixed state that is reminiscent of other ultra-clean nodal superconductors.

\section{Results and Discussion}

\begin{figure}
\includegraphics[width=8.5cm]{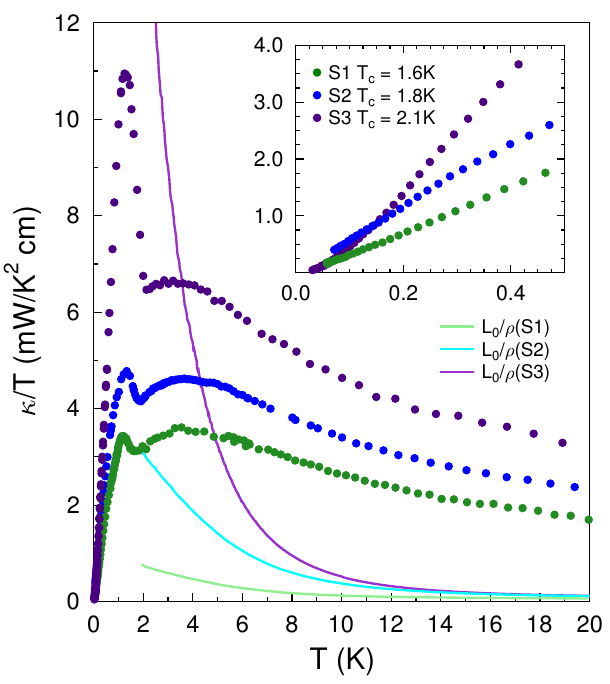} 
\caption{{\bf Variation of the {$a$-axis} thermal conductivity of  UTe$_2$ with disorder.} 
The measured thermal conductivity of three samples of UTe$_2$ with varying residual resistance ratios (RRRs) of 30 (S1, green), 50 (S2, blue) and 600 (S3, purple) and superconducting transition temperatures of 1.6~K (S1), 1.8~K (S2) and 2.1~K (S3) are presented as a function of temperature. The overall shape of the curves is similar for all three samples, but with noticeably stronger peaks below $T_c$ in the higher RRR samples. The estimated electron contribution to the total thermal conductivity estimated by the Weidemann-Franz law ($L_0/\rho(T)$) is shown for each sample as a thin line. The inset shows the measured thermal conductivity data below 0.5~K, with clear limiting behavior approaching $\kappa/T$=0 as $T\to 0$ for all samples.}
\label{fig:Kappa_v_T} 
\end{figure}

Data are presented on four single-crystal samples of UTe$_2$ with varying T$_c$ values and normal state scattering rates as deduced from their residual resistivity ratios (RRR = $\rho(300~$K$)/\rho(0~$K$)$). Sample S1 ($T_c$=1.6~K, RRR=30), grown by the chemical vapor transport (CVT) method and whose thermal conductivity was previously reported \cite{metz_point-node_2019}, is compared to sample S2 ($T_c$ =1.8~K,RRR=50) grown by CVT at lower temperatures (950C and 860C), and  samples S3 ($T_c$ =2.1~K,RRR=600) and S4 ($T_c$ =2.1~K, RRR=100) grown by a novel tellurium flux method described in the Materials and Methods section.

Electrical and thermal transport of samples S1, S2, and S3  are presented in Figure \ref{fig:Kappa_v_T}, plotted as the total measured thermal conductivity $\kappa/T(T)$ and the electron contribution $L_0/\rho(T)$ for each sample estimated from the Wiedemann-Franz law, $\kappa/T=L_0/\sigma$ (all for currents directed along the crystallographic $a-$axis).
In the normal state, all three samples show qualitatively similar behavior as a function of temperature, albeit with larger absolute values of $\kappa/T$ in the samples with lower residual resistivities. 
Below $T_c$, all three samples exhibit a significant increase in $\kappa/T$ that peaks near $\sim$1~K but with a peak magnitude that varies considerably between the samples, being the most pronounced in S3 followed by smaller magnitudes for S2 and S1, respectively. 
The magnitude of this enhancement varies inversely with the residual resistivity of each sample, also reflected in the RRR values (RRR= 30, 50, and 600 for samples S1, S2 ans S3 respectively).
As explained below (see "Theorretical Analysis" Section), this behavior is qualitatively consistent with that of many unconventional superconductors, which frequently exhibit such enhancements in $\kappa/T$ below the superconducting transition that scale inversely with the elastic scattering rate. 

%\onecolumngrid

\begin{figure*}
\includegraphics[width=17.5cm]{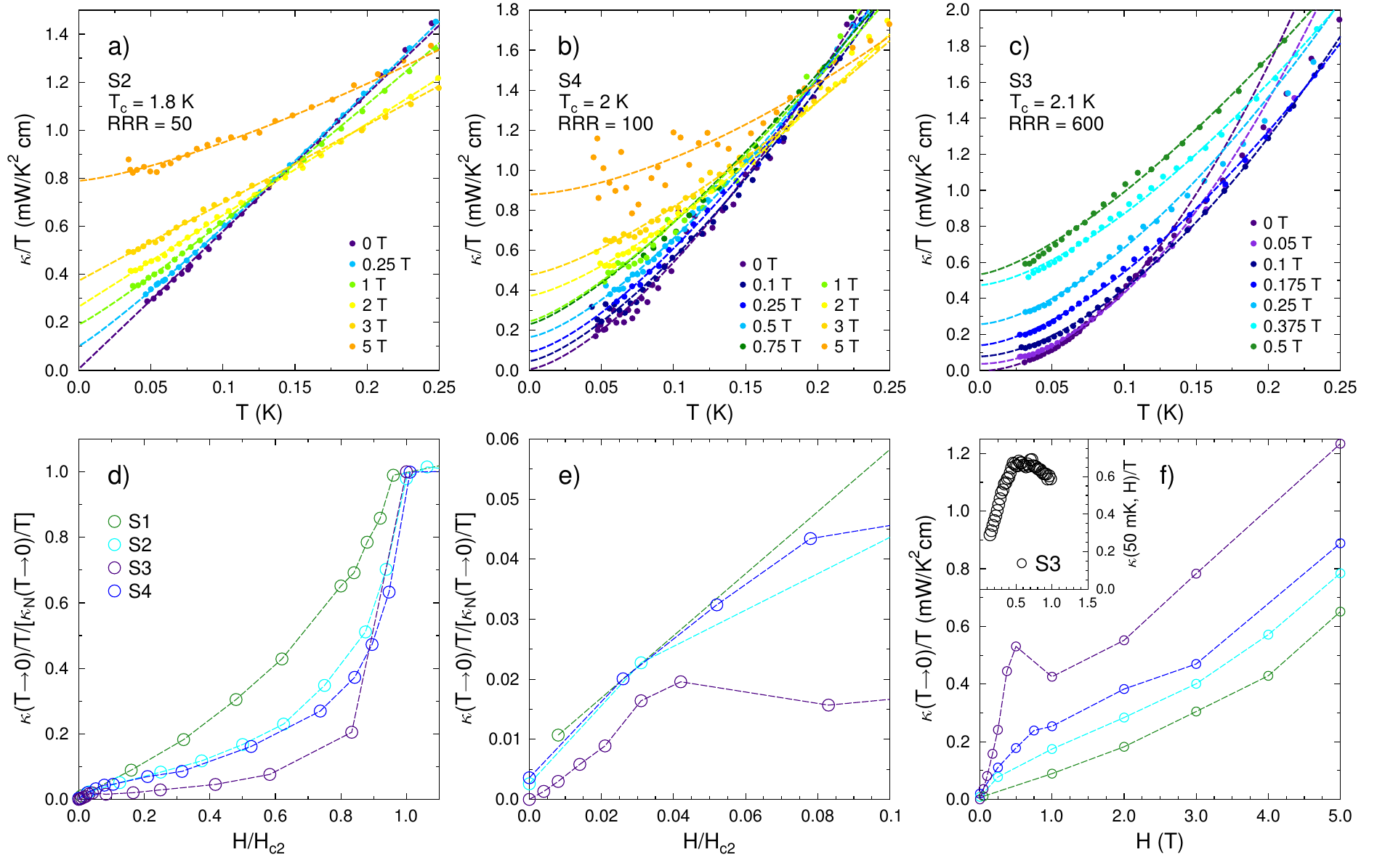} 
\caption{{\bf Thermal transport of UTe$_2$ in the superconducting state.} 
%{\red [FIX Tc VALUES - b) 2.1K, c) 2.1K?; also fix order S1 S2 S3 S4 in legends d-f]} 
Data are presented for four comparative samples with varying $T_c$ and RRR values as indicated.
Panels (a), (b), and (c) present the magnetic field evolution of $\kappa(T)/T$ at low temperatures for the new samples measured in this study. Qualitatively similar behavior is observed in all samples, with zero-field data trending clearly toward zero conductivity values as $T \to 0$ for all samples, and rapid increases of the extrapolated value with increasing magnetic fields.
The extrapolated $T \to 0$ values for each curve are plotted as a function of magnetic field $H$ in panels (d), (e) and (f). 
Panel (d) plots the full field range up to the upper critical field $H_{c2}$, with data normalized to its normal state value above $H_{c2}$ and plotted against normalized field $H/H_{c2}$, showing the apparent depression of conductivity with increasing RRR value of sample. 
Panel (e) presents the same normalized data as panel (a) but zoomed in to the very low  field range where the universal rapid initial rise is evident.
Panel (f) presents the same data but plotted as a function of absolute conductivity and magnetic field values, demonstrating the enhancement of absolute conductivity with increasing RRR and $T_c$ values, but also the remarkably consistent rate of growth above 1~T fields as discussed in the text. The inset to Panel (f) shows a field sweep of sample S3 at 50mK, demonstrating that the peak shown in the zero temperature extrapolation is not an artifact. }
\label{fig:Kappa_v_H} 
\end{figure*}
%\twocolumngrid
 
%kappa/H discussion
As discussed below, the enhanced thermal conductivity below $T_c$ can be naturally explained using model calculations that vary the electron scattering rate, but these models also require a mixture of electron and phonon contributions.
To start, we focus on the most reliable method of measuring the electronic response due to the superconducting gap structure, which is to study the evolution of the zero-temperature extrapolation of $\kappa/T$ with magnetic field. In this limit phonons necessarily vanish and the behavior of the extrapolated value, $\kappa(T \rightarrow 0, H)/T$, can only reflect the properties of the superconducting state. 
In this regime, the change of the thermal conductivity under a small magnetic field is dominated by a shift in the quasiparticle spectrum that is induced by the presence of the supercurrent associated with vortices.\cite{Shakeripour_heat_2009} If the quasiparticle spectrum has a gap, then a nonzero magnetic field is required before $\kappa/T$ exhibits any increase. In contrast, if there are nodes in the spectrum then at arbitrarily small fields the density of states will be enhanced and $\kappa/T$ will increase continuously from zero field. This effect has been used to diagnose the presence of nodes in the gap structure of many superconductors \cite{Shakeripour_heat_2009}, including in the earliest work on UTe$_2$ \cite{metz_point-node_2019}.

Figure \ref{fig:Kappa_v_H} presents the temperature and magnetic field dependences of the fermionic thermal conductivity, $\kappa(T, H)/T$, for samples S2, S3, and S4, plotted as a function of temperature for various fixed fields ($\kappa(T, H)/T$, panels a-c), as well as a function of field in the $T\to 0$ limit ($\kappa_0(H)/T$, panels d-f).
In all of these measurements, the magnetic field was applied in the longitudinal configuration, with field parallel to both the heat current and the crystallographic $a-$axis. The overall shape of  
$\kappa(T, H)/T$ is qualitatively similar for all  samples, with a rapid increase with temperature that is mildly dependent on magnetic field strength at low fields, but with a residual $T\to 0$ intercept $\kappa_0/T$ that exhibits a strong field dependence.
As shown in Figs.~\ref{fig:Kappa_v_H}d)-f), 
$\kappa_0/T$ undergoes an initial rapid rise as the field is increased from zero, followed by an intermediate field regime where it grows more slowly. Once the magnetic field is increased to nearly $H_{c2}$, $\kappa_0/T$ rises rapidly to the normal state value. We study the low field regime most fully in the highest RRR sample, S3, where data were taken down to 30~mK for six magnetic fields up to 0.5~T provides the most robust $T\to 0$ extrapolations (Fig.~\ref{fig:Kappa_v_H}c). $\kappa(T)/T$ exhibits a clear, rigid upshift with increasing fields starting as low as 50~mT. We emphasize that this rise at low fields is readily visible in our raw data at the lowest temperatures and is therefore not sensitive to the details of the $T\rightarrow 0$ extrapolation. However, because the temperature growth in $\kappa/T$ is stronger than linear, this extrapolation can be misleading if data is not measured below $\sim$100~mK, as has been reported recently.\cite{suetsugu2023}

{The idea that this low field rise is coming from a Doppler-shifted density of states is consistent with the variation in the magnitude of this effect in these samples. To factor out changes coming from variation in the mean free path of the different samples we can use the zero temperature value of $\kappa/T$ in the normal state above $H_{c2}$. In the normal state all of these samples should have the same carrier density, so changes in $\kappa_0/T$ should scale with the mean free path. In Fig \ref{fig:Kappa_v_H}e) we plot $\kappa(T \rightarrow 0, H)/\kappa(T \rightarrow 0, H_{c2})$ as a function of normalized magnetic field, $H/H_{c2}$. All for samples show a comparable slope, suggesting that they have a common density of states that is being tuned by the magnetic field.}

Further support for the Doppler shift interpretation comes from the fact that $\kappa/T$ rises immediately starting at zero field.
There are various scenarios in which a fully gapped superconductor can exhibit a small rise in $\kappa/T$ at magnetic field values well below $H_{c2}$, including the presence of very deep gap minima and multi-gap scenarios. However, in any scenario with a full gap there will be a finite field range within which $\kappa/T$ is approximately zero, with the rise in $\kappa/T$ accelerating only as the magnetic field overcomes the minimum of the gap, leading to a concave up shape for $\kappa_0(H)/T$. 
By contrast, the extrapolated value of $\kappa(T, H)/T$ grows most quickly near zero field for all of our samples, allowing us to conclusively rule out a fully gapped superconducting state in UTe$_2$ at low pressures and magnetic fields.

Having concluded that UTe$_2$ poses nodes, the next question is whether these are point nodes or line nodes.
There is a well-known theoretical result that in a superconductor with line nodes the residual value of $\kappa/T$ in the zero temperature limit is bounded from below.\cite{Graf1996}
This has been found to be the case empirically in several line-node superconductors including KFe$_2$As$_2$ and YBCO \cite{Reid_universal_2012}. 
This universal residual term is given by $\kappa_{0}/T=\frac{1}{3}\gamma_0 v_f^2 \frac{a\hbar}{2\mu \Delta_0}$ \cite{Graf1996}. Using the best estimates from the literature for $2\Delta_0$=$3.5k_BT$, $\gamma_0$=125~mJ/mol K$^2$, $\mu =2$ and $v_f =10,000$~m/s, this yields a residual term of about 0.25~mW/K$^2$cm.
As shown in the Fig.~\ref{fig:Kappa_v_T} inset, the zero-field, ultra-low temperature behavior of $\kappa/T$ for samples S1, S2, and S3 clearly extrapolates to well below 0.25~mW/K$^2$cm at $T$=0, violating the minimum value expected for a line node superconductor. 
Together with the evidence for a rapid rise in 
$\kappa_0/T$ at low fields, this provides conclusive evidence for symmetry imposed point nodes in the gap structure of UTe$_2$.
This is consistent with conclusions from nearly all thermodynamic experiments to date  -- including  specific heat \cite{ran_nearly_2019,Rosa2022,Kittaka}, penetration depth \cite{metz_point-node_2019,Ishihara2023} and NMR \cite{Nakamine2021, nakamine2019} -- and, together with the theoretical analysis presented below, point to a gap structure consistent with point nodes lying away from the crystallographic $a$-axis.

Finally, we identify some intriguing universal behavior of $\kappa_0(H)/T$ in the intermediate field regime.
As discussed above and shown in Fig \ref{fig:Kappa_v_H} panels e) and f), the higher RRR samples show a more rapid rise in $\kappa/T$ with applied magnetic field. However this trend is confined to a region below 1~T. At higher fields, the unnormalized $\kappa(T \rightarrow 0, H)/T$ grows at a nearly uniform  rate in all samples, despite the factor of $\sim 20$ difference in their residual resistivities. 
%{\blue [maybe better to say 'equal to within xx \% of the normalization factor used above']}
As the disorder is reduced, the conductivity is simply increased by a fixed amount independent of field. Generically, the physics of the mixed state is expected to be quite complicated since it involves quasiparticle dynamics in the inhomogeneous medium of a vortex lattice, as well as the dynamics of the vortices themselves. The fact that disorder influences this regime in an extremely simple way in UTe$_2$ strongly suggests that there is a simple principle or physical picture that determines $\kappa/T$ in this regime. This potentially opens up the mixed state to more thorough theoretical treatment than has been attempted previously. 

The most important observation to be made about the data in this regime is that this simple pattern exists in $\kappa(T \rightarrow 0, H)/T$ as a function of disorder. However, we offer a few thoughts about what it might imply for the physics of the mixed state.
First of all, this separation -- between an overall conductivity scale that is sensitive to disorder, and a field dependence that is not -- is suggestive of parallel conductivity channels. If a resistivity shows a simple offset between samples with different disorder levels, it is easy to interpret that behavior as arising from the presence of multiple scattering rates. Given that this pattern arises in the conductivity, it is more easily explained as arising from the presence of two conductivity channels, one of which is field independent and affected by disorder, and one of which sensitive to magnetic field but insensitive to disorder.
Second, the fact that the boundary between these two regimes lies at the same value of the magnetic field in all samples suggests that this is not a vortex-lattice transition. Such a transition should depend on the density of pinning sites and therefore is expected to be disorder-dependent. 
Third, a fixed field scale translates to a specific vortex density, so it seems likely that this high field regime is defined by some length scale in the system becoming comparable to the inter-vortex distance.

One consequence of this disorder-independence of the rise of conductivity with field is that the shape of  $\kappa_0(H)/T$ is flatter in cleaner samples when plotted as a fraction of the normal state thermal conductivity (i.e. Fig.~2d ). This trend has been noted in several very clean unconventional superconductors such as KFe$_2$As$_2$ and YBCO.\cite{Reid_universal_2012} To the best of our knowledge, none of the data sets on these materials presents a wide enough array of disorder values to thoroughly test whether there is a disorder-independent growth rate for $\kappa/T(H)$ in any of these materials. If it is the case that any of these other clean nodal superconductors show similar behavior as a function of disorder, it would create a basis for understanding the mixed state that is independent of the details of pairing.
This creates a powerful motivation for more thorough studies of the mixed state in other clean nodal superconductors.   

Finally, there is the striking observation that for sample S3, which has the highest RRR value, $\kappa(T \rightarrow 0, H)/T$ actually reaches a maximum at 0.5~T before decreasing and settling into a smooth rise in the intermediate field regime. While it may be tempting to interpret this local maximum as an indication that some distinctive physics is present around 0.5~T, such as a phase transition, the simplest  interpretation is probably that this peak is just a consequence of the crossover from the low-field regime dominated by the Doppler shift of the quasi-particle spectrum to the intermediate field regime which is dominated by some other physics. All of our samples exhibit a crossover at this absolute field scale, so it is reasonable that in the highest RRR sample $\kappa_0(H)/T$ grows fast enough that it overshoots the value that it will ultimately settle into in the universal mixed regime above 1~T.
%Further experiments and modeling will be required to know for sure why $\kappa(H)/T$ exhibits a peak at low levels of disorder. {\blue [last sentence can be x'd]} 

%At the present time there is no simple way to explain or model these patterns, but they should provide valuable basis for future studies.

%Furthermore, the trend in the absolute values of $\kappa/T$ in these samples supports
%The above data clearly support the presence of nodes in the superconducting gap of UTe$_2$. The next question is whether these are point nodes or line nodes. Strong evidence in favor of the point node scenario can be found in the  zero temperature, zero field behavior of $\kappa/T$. There is a well-known theoretical result that in a superconductor with line nodes the residual value of $\kappa/T$ in the zero temperature limit is bounded from below. This has been found to be the case empirically in several line-node superconductors including KFe$_2$As$_2$ and YBCO\cite{Reid_universal_2012}. Figure \ref{fig:Kappa_v_T} (B) shows the very low temperature behavior of samples S1, S2, and S3 in zero magnetic field. For every sample, $\kappa/T$ clearly extrapolates to zero at zero temperature, violating the minimum value expected if there were line nodes. It thus seems clear that the nodes in UTe$_2$ have to be point nodes.

\begin{figure}
\includegraphics[width=1\linewidth]{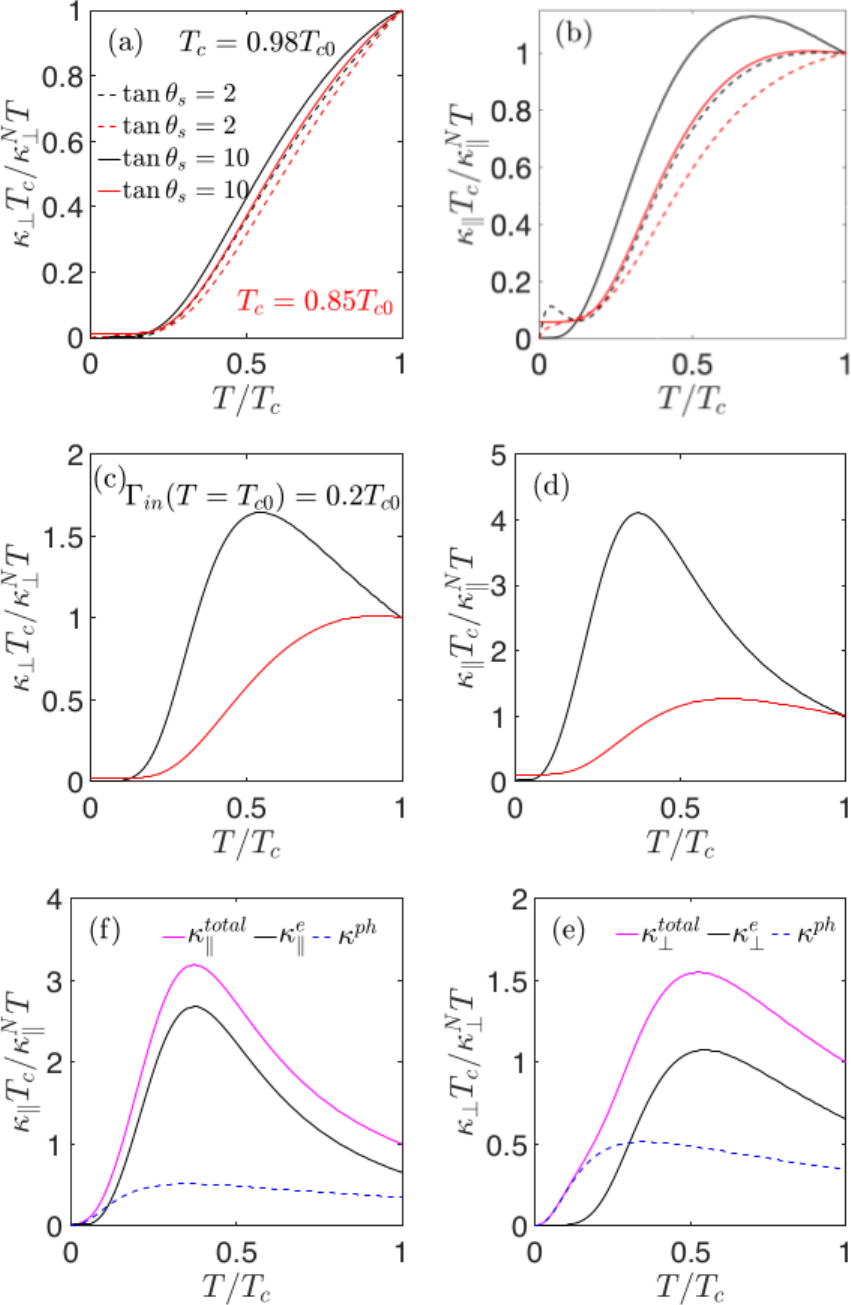} 
\caption{{\bf Thermal conductivity for an axial state on a spherical Fermi surface.} 
The calculated thermal conductivity is presented for a gap model with $\Delta=\Delta_0 \sin \theta \exp(i \phi)$, where $\theta$ is the polar angle and $\phi$ is the azimuthal angle, for heat currents directed perpendicular (left columns) and parallel (right columns) to the nodal positions, and for two different disorder levels (0.98$T_{c0}$ in black, 0.85$T_{c0}$ in red).
Panels (a) and (b) present electronic thermal conductivity as functions of temperature for weak ($\tan \theta_s=2$, dashed lines) and for the strong ($\tan \theta_s=10$, solid lines) pointlike impurity scatterers.
Panels (c) and (d) include the contribution from the inelastic scattering to the electronic thermal conductivities in the presence of strong scatterers with $\tan \theta_s =10 $, and panels (e) and (f) add phonon contributions (blue dashed lines) modeled assuming defect scattering as described in the text. The total thermal conductivity (magenta solid lines) thus is calculated to include impurity scattering with $\tan \theta_s =10 $ for $T_c=0.98T_{c0}$, inelastic scattering $\Gamma_{in}=0.2T_{c0}$ at $T=T_{c0}$ and phonon contributions.}
\label{fig:Theory1} 
\end{figure}

%[START THEORY SECTION]

\section{Theoretical analysis}
Here we validate the point node picture quantitatively with a model calculation of the temperature dependence of $\kappa/T$.
The theory of thermal conductivity  in triplet superconductors is very similar to that for singlet superconductors\cite{AmbegaokarGriffin:1994,KadanoffMartin1963}, provided the triplet order parameter matrix is unitary.  Most of the commonly studied triplet states, including the $^3$He-A phase, are in this class. In that case the familiar quasiparticle energies $E_\k=\sqrt{{\xi_\k}^2+|{\bf d}(\k)|^2}$ enter the thermal current response, and the same expressions can be used for triplet superconductors\cite{SSchmitt-Rink:1986,SSchmitt-Rink:1986,PJHirschfeld:1986,PJHirschfeld:1988}.  In the case of nonunitary states, additional terms involving ${\bf q}\equiv i{\bf d}({\bf k}) \times {\bf d}^*({\bf k})$  occur in both the quasiparticle energies and in the weights of two-particle processes\cite{MishraHirschfeld2023}.  In the nonunitary case, the zeros of $|{\bf d}({\bf k})|^2$ differ from those of the zeros of $E_{\k}$ even on the Fermi surface $\xi_{\k}=0$ (``spectral nodes"). Indeed, it was proposed by  Ishihara et al.\cite{Ishihara2023} that in UTe$_2$, complex linear combinations of 1D irreducible representations could  support spectral nodes pointing in generic directions in the orthorhombic Brillouin zone,  explaining early experiments indicating time reversal symmetry breaking \cite{Hayes2021}.  By contrast, order parameters corresponding to a single 1D irreducible representation are necessarily real, with nodes aligned along high-symmetry axes.  Previous work measured penetration depth for currents running along different directions, and concluded from the universal $\delta \lambda(T)\sim T^2$ behavior that nodes were located at generic positions in the Brillouin zone, requiring nonunitary states\cite{Ishihara2023}. 

%While some directional information is available for the thermal conductivity of earlier generation samples, 
Thermal conductivity is complex to analyze due to the fact that thermal currents can be carried by both electrons and phonons, each of which can scatter from multiple physical sources.  In cuprates, it was shown that electronic conductivity limited by disorder was dominant at low temperatures, while a combination of inelastic electron-electron scattering and phonon-electron scattering dominated
the intermediate and high temperature regimes\cite{Hirschfeld1996}.  Intriguingly, the directional thermal conductivity of sample S1  \cite{metz_point-node_2019} is unexpectedly relatively isotropic, at least in the context of a system with point nodes along a single high-symmetry axis.  This could reflect a distribution of nodes away from high symmetry directions, as suggested in Ref. \cite{Ishihara2023}, or a larger than expected phonon contribution.  
We note, for instance, that the low temperature thermal conductivity of samples S1, S2 and S3 are not strikingly different as shown in Figure 1 (inset), suggesting that phonon conduction is not negligible. 
 Theoretically, calculated values of electronic $\kappa/T$ are too small at low $T$ and fall systematically below the data, providing further support for this conclusion.

\begin{figure*}
\includegraphics[width=1\linewidth]{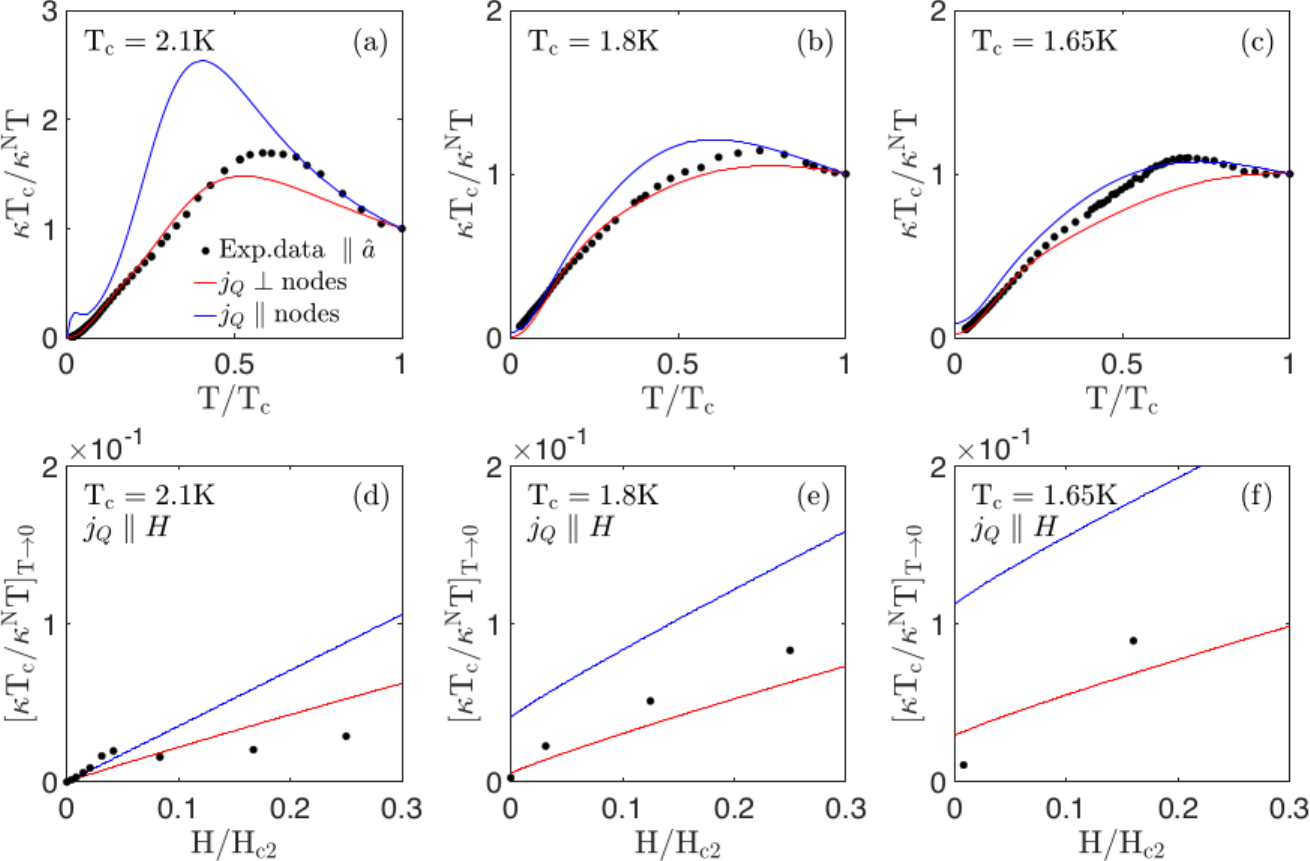} 
\caption{{\bf Qualitative fits to the experimental data . } Panels (a)-(c) show the qualitative fit to the temperature dependence of   $\kappa/T$ normalized to their respective normal state values  $\kappa_{\mathrm{N}}/\mathrm{T}$ at $T_c$. The $s$-wave scattering phase-shift is $\tan^{-1}3$, and $\kappa/T$ along the nodal direction (blue) and perpendicular to the nodal direction (red) include contribution from phonons and inelastic scattering. Filled circles show the experimental data for heat current applied along the $a$-axis. Panels (d)-(f) show the fits (lines) to the field dependence of the zero temperature limit values of experimental data (circles) measured with heat current parallel to applied fields, normalized to the zero temperature limit values of $\kappa/T$ measured at $H_{c2}$. Legends are equivalent for all panels and each column represents same values of $T_c$ and impurity parameters. 
Note that the impurity scattering rates are chosen to match the experimental $T_c$ values.}
\label{fig:Theory2} 
\end{figure*}

We assume the standard Bardeen-Rickayzen-Tewordt theory of phonon thermal conductivity\cite{BRT1959,Uher1990} with phonon mean free path limited by defects.
For simplicity, we ask initially  what information can be gleaned from the qualitative features of the temperature- and field-dependence of the measured thermal conductivity along the $a$-axis direction only.   The enhancement of the electronic mean free path below $T_c$ due to the collapse of inelastic scattering, followed by a saturation of the mean free path due to disorder,   led in the cuprates to a peak in $\kappa_{el}$ that was very sensitive to disorder.  This behavior is similar to that observed  in Fig. \ref{fig:Kappa_v_T} for our UTe$_2$ samples, so we attempt a similar analysis here, with  phonon contributions $\kappa_{ph}$ and inelastic scattering due to electronic interactions $1/\tau_{inel}$  that by assumption do not depend on pairbreaking  disorder.  The electronic part $\kappa_{el}$ is calculated assuming independent elastic and inelastic contributions, $1/\tau = 1/\tau_{el}+1/\tau_{inel}$,   and added to the phonon contribution.  The model and parameters used are described in the Appendix.  

In Fig. \ref{fig:Theory1}, we show the theoretical thermal conductivity $\kappa(T)/T$ calculated with these assumptions in an axial state, ${\bf d}(\k)=(k_x+ik_y){\hat y}\propto \sin\theta$, with $\theta$ the polar angle with respect to the $z$-axis in momentum space, taken to lie along the crystallographic $a$-axis direction. This choice of order parameter is obviously a dramatic oversimplication compared to the states that have been discussed previously in the orthorhombic crystal, including nonunitary ones; nevertheless the qualitative features at low $T,H$ should depend only on the existence of linear point nodes, and their location relative to the field and current directions. Panels (a),(b) of Fig. \ref{fig:Theory1} show some representative electronic thermal conductivity plots assuming scattering from impurities only, for a range of impurity parameters; the plot is based on the theory reviewed in Appendix A.  First, note that the observed $\kappa_{el}/T\sim T^2$ behavior is expected over an intermediate temperature range for a state with point nodes located in a generic position relative to the heat current flow, and impurity scattering in the intermediate-to-unitary limit.  At asymptotically low temperatures, the $T$ dependence is strongly influenced by self-consistency, despite the lack of leading universal constant $\kappa_{el}/T$ as in the line node case\cite{Graf1996,Norman1996}.  Instead, the behavior of $\kappa_{el}/T$ as $T\rightarrow 0$ is very flat\cite{Hirschfeld1986,Hirschfeld1988}.  Furthermore, as shown in Fig. \ref{fig:Theory1}(b), deviations from  unitary or near-unitary scattering lead to poor agreement with the data due to low-$T$ upturns.

Fig. \ref{fig:Theory1}(c) and (d) now illustrate the effect of inelastic scattering on the electronic thermal conductivity.  The strong peak below $T_c$ in $\kappa/T$ is strongly reminiscent of cuprates, where it is attributed, e.g. in YBCO\cite{Hirschfeld1996}, to the collapse of the quasiparticle spin fluctuation relaxation rate in the superconducting state as the gap opens below $T_c$.   In the current model, however, the collapse is somewhat weaker due to the point nodes, and the contribution of the inelastic scattering falls off even more rapidly than in the line node case as $T\rightarrow 0$ (Appendix A).  In Fig. \ref{fig:Theory1} (e),(f), we show the effect of adding the phonon contribution to the thermal conductivity.  Phonons are modeled assuming defect scattering as described in Refs. \cite{Uher1990,Hirschfeld1996} and reviewed in Appendix A.  Within the current model, this leads to $\kappa/T=\kappa_{el}/T+\kappa_{ph}/T$ peaks at somewhat lower temperatures than experiment, but agrees qualitatively with the overall behavior with temperature (Fig. \ref{fig:Theory2}).

The  experimental data shown in  Fig. \ref{fig:Kappa_v_T} vary monotonically at low $T$, and there is no linear-$T$ term in $\kappa$ from impurity scattering, as expected for an axial state realized in sufficiently clean samples\cite{Graf1996,Norman1996}.  We show in Fig. \ref{fig:Theory2}(a) a comparison of the best fit of the theory for a combination of impurity, inelastic electron scattering and phonon scattering parameters, with the experimental data. 
Since  the variation of physical properties with sample quality is a pressing issue in UTe$_2$, we also  present  results in Fig.\ref{fig:Theory2}(a)-(c) for concentrations corresponding to $T_c$ values of 2.1 K, 1.8 K and 1.6 K in theory and experiment.   We see that the  height of the inelastic peak at intermediate temperatures  grows with decreasing disorder, as  in Fig. \ref{fig:Kappa_v_T}, reflecting the growth of the inelastic quasiparticle mean free path to much lower temperatures before being cut off by the elastic mean free path.
While many uncertainties exist with this simplified model as discussed below, the best fit is obtained assuming the nodes are away from the $a$-axis, most readily observed in the large calculated anisotropy for node-parallel and -perpendicular heat currents in Fig.~4a). In part this conclusion arises from the sensitivity of the $j_q\parallel \hat a$ result to impurity phase shift deviations from unitarity, as reflected in the low-$T$ bump not seen in experiment. But it is also consistent with recent experiments suggesting the nodal positions are located away from the $a$-axis, in particular the recent ultrasound experiments that find less sensitivity to strain along the $b$-axis as evidence for a B$_{2u}$ order parameter \cite{theuss2023}, as well as a small anisotropy in penetration depth consistent with $b$- and $c$-axis nodal directions \cite{Ishihara2023}.

We now compare with the expected field dependence for a state with point nodes, calculated within the Doppler shift approximation where quasiparticle motion in the vortex state is treated semiclassically, such that each quasiparticle has its energy shifted $E_{\bf k}\rightarrow {\bf v}_s\cdot {\bf k}$, where ${\bf v}_s$ is the superfluid velocity. The Doppler shift contribution from extended quasiparticle states in this case can be shown to be $\kappa/T(H\rightarrow 0)\sim H/H_{c2}$, very different from the exponentially activated dependence expected for a fully gapped system, and somewhat different from the $H \log H$ behavior expected for line nodes.   The vortex lattice itself is assumed to be pinned (with other assumptions and parameters  described in Appendix A).    In Fig. \ref{fig:Theory2} (d)-(f), we show the calculated field dependent $\kappa$, normalized to its $\Delta=0$ (normal state) 
$T\rightarrow 0$ value, which in the current description corresponds to the electronic impurity scattering contribution only.  This is the same procedure employed in the normalization of the experimental data in Fig. \ref{fig:Kappa_v_H}(d),(f).  While the theory does not reproduce the apparent small peak at low fields for the purest sample, it is roughly consistent in magnitude and behavior with the overall field behavior for $H\ll H_{c2}$.  Thus, while the best fit for  sample S3 (Fig. \ref{fig:Theory2}(d)) is consistent with the conclusion drawn from the $T$-dependence, that nodes are away from the principal $a$-axis direction, the results for the dirtier samples, S1 and S2 in (e),(f), appear to disagree. However, the rough suppression of the overall field dependence variation with disorder is captured by the theory.   We note that this magnitude of the field variation includes a factor describing the geometry of the vortex lattice and the coherence lengths; this factor is not known quantitatively but should be of order unity and is chosen to be the same for all three disorder levels.  Thus from the field dependence alone it is  difficult to determine the positions of the nodes relative to the heat current. On the other hand, it is clear that the field dependence is most consistent with point nodes, and inconsistent a fully gapped state.

\section{Conclusions}
In this work,  we report the first systematic study of the low-temperature thermal transport of UTe$_2$ as a function of scattering rate and superconducting transition temperature, utilizing crystals obtained by different growth methods including a unique synthesis route using tellurium flux to obtain residual resistance ratios as high as 600.  
Given the variations in sample properties reported for the evolving generations of crystalline quality, the direct measure of the density of low-energy quasiparticle excitations this experiment provides yields conclusive information about the intrinsic nature of the superconducting gap symmetry and structure.
We have measured the temperature- and magnetic-field dependence of the thermal conductivity of four samples with varying $T_c$ values and normal state scattering rates, revealing a robust pattern of behavior in the superconducting state of UTe$_2$. Our experimental data can be understood within the standard theory of a superconductor with point nodes in the presence of varying levels of disorder, including inelastic electron-electron scattering that leads to a peak in the temperature dependence below $T_c$ for the cleaner samples.  The zero-temperature and low-field limiting behaviors of thermal transport are shown to only be consistent with the presence of a superconducting gap with point nodes, allowing us to rule out any full-gap scenarios. Theoretical analysis of a simple model $p$-wave state supports a picture with nodes located away from the crystalline $a$-axis, suggesting a B$_{1u}$ or B$_{2u}$ symmetry for the low-field, ambient-pressure superconducting order parameter of UTe$_2$ that is consistent with recent experiments. The sample-independent rate of increase of thermal conductivity observed in the zero-temperature, intermediate-field mixed state is suggestive of a universal phenomenology for ultra-clean nodal superconductors that deserves further attention.

{\bf Acknowledgements.}  The authors acknowledge useful conversations with D. Agterberg, J.C. Davis, V. Madhavan, and S. Anlage.  
Research at the University of Maryland was supported by the Department of Energy Award No. DE-SC-0019154 (transport experiments), the Gordon and Betty Moore Foundation’s EPiQS Initiative through Grant No. GBMF9071 (materials synthesis), NIST, and the Maryland Quantum Materials Center.
PJH was supported by NSF-DMR-2231821.

%In this work we have pushed the study of the thermal conductivity of UTe$_2$ to the most extreme combination o of low temperature and high conductivity yet achieved. Despite wide variation in sample conductivity, we observe a universal pattern of behavior in the variation of $\kappa/T$ at low temperatures as a function both of temperature and magnetic field, a pattern which is only naturally compatible with the presence of point nodes in the gap structure. This universal nodal behavior is strong confirmation of the triplet nature of the order parameter and provides some important constraints on which irreducible representation of the crystal point group the orbital part of the order parameter belongs to. Additionally, we report the first synthesis of superconducting samples of UTe$_2$ from tellurium flux. This sudden appearance of superconducting crystals in growths with 32$\%$ or more uranium is intriguing and points to an essential role of $U_7Te_12$ in the growth of very high conductivity samples. Further work will be required to illuminate that role, as well as to make sense of the behavior of $\kappa/T(T=0)$ in the intermediate field regime, particularly the peak observed in highest RRR sample around 0.5 tesla and the quasi-universal rate of growth of $\kappa/T$ in this regime.

\clearpage
\setcounter{figure}{0}
\renewcommand{\thefigure}{A\arabic{figure}}%
\appendix

\section{ Materials and Methods.}
%materials and methods
    %Discuss CVT and Flux growth, 
    Two different growth methods were employed to produce crystals with reduced normal state scattering rate. The first was a chemical vapor transport method analogous to the one used in ref \cite{metz_point-node_2019}, but at reduced temperatures. This was motivated by previous work showing that lower growth temperatures can increase $T_c$ and reduce normal state scattering in UTe$_2$ \cite{Rosa2022}. The sample measured here was grown for one week in a temperature gradient of 950C to 860C, with a uranium to tellurium ratio of 1:1.85. Samples from this growth batch consistently exhibited superconducting transitions at 1.8 kelvin. The residual resistance ratios of crystals in this batch varied from 30-50. 
    %A sample with RRR = 50 was selected for thermal conductivity measurements, and its resistance versus temperature curve is plotted in Figure .

    Further increases in $T_c$ were obtained through a novel tellurium-flux growth method. Previous attempts to grow UTe$_2$ in a tellurium flux have only yielded non-superconducting crystals with very low residual resistivity ratios of about 2 or three. The best available analyses of these samples suggest that they are uranium deficient at the few percent level \cite{Sakai2022}. Motivated by these observations, we attempted Te-flux growths with greater concentrations of uranium. For these growths, uranium and tellurium chunks were combined in a ratio of nearly one to two, placed inside an alumina crucible and sealed in a quartz ampoule with an argon atmosphere of about 100mbar. The growths were heated rapidly to 1185C, held there for five hours and then cooled to 1000C at 4C/hour. Because the uranium to tellurium ratio was close to the stoichiometry of the UTe$_2$, relatively little excess tellurium remained after the growth cycle. This allowed the ampoule to be simply cooled to room temperature and the residual flux was removed from the crystals mechanically.

\begin{figure}
\includegraphics[width=8.5cm]{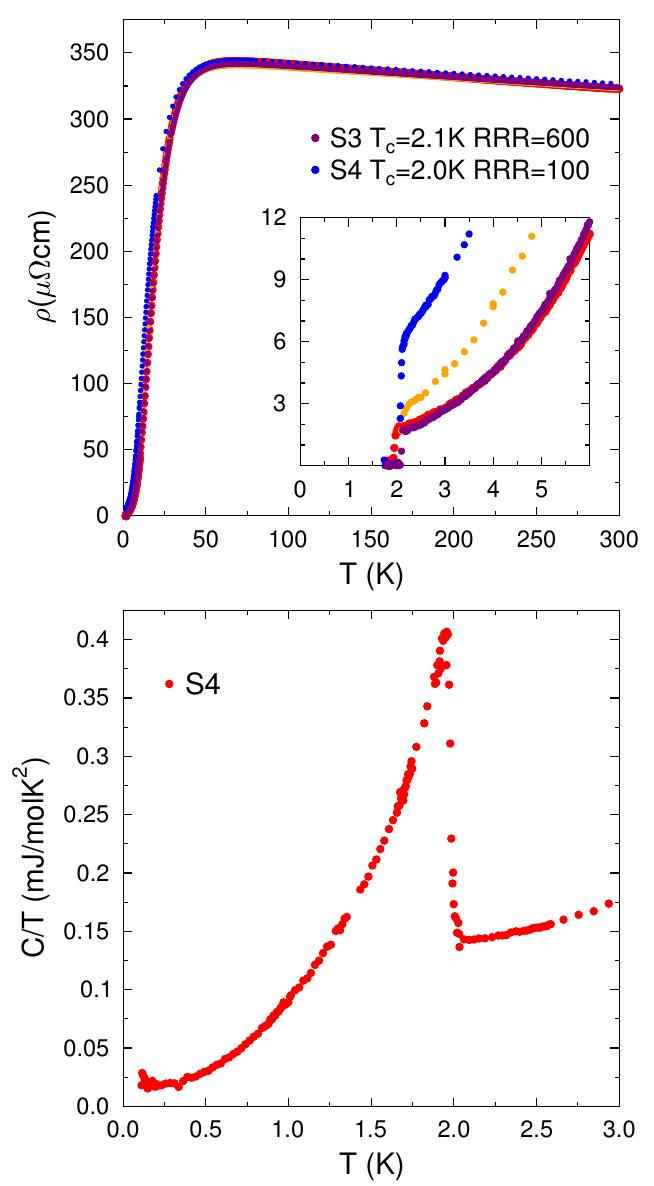} 
\caption{{\bf Characterization of superconducting UTe$_2$ samples grown in Tellurium flux.} Panel (a) shows the $a-$axis resistivity of four samples of Te-flux grown UTe$_2$. All four samples show a superconducting transition around 2$~$K and RRRs above $\sim$100. The data are normalized to the average value of $\rho(300~K)$ for consistency. Panel (b) shows the superconducting transition in heat capacity of S3 from the main text. The size of the jump at T$_c$ is comparable to that of high-T$_c$ samples of UTe$_2$ grown by other methods.}
\label{fig:R_and_C} 
\end{figure}
    
    In growths with $\sim$30$\%$ uranium content, the RRRs of the samples improved to about seven--as opposed to two or three as found in samples grown with an initial uranium concentration of 22$\%$ or 25$\%$\cite{Sakai2022}. At a 32$\%$ uranium concentration, the growths produced some superconducting crystals. These crystals make up about a third of the the overall yield of the growths\textemdash with the remainder being nonsuperconducting UTe$_2$ crystals\textemdash and they exhibit superconducting transition temperatures of about 2K and elevated residual resistivity ratios of between 100 and 600. By contrast, the non-superconducting crystals in these growths show RRRs of about 3. These observations suggest chemical similarity to crystals grown in a NaCl/KCl flux, which also exhibit superconductivity at 2K and arise characteristically as a minority phase in those growths\cite{Sakai2022}. Further work will be required to determine exactly why superconductivity emerges suddenly in growths with a 32$\%$ uranium concentration, but we can offer a few observations here. According to published binary phase diagrams, a Te-U system with 32$\%$ or more uranium will first form the phase U$_7$Te$_{12}$ as it cools and then form UTe$_2$ crystals. The fact that superconductivity is not observed in any crystals grown at 30$\%$ or less strongly suggests that the presence of U$_7$Te$_{12}$ in the melt is at least part of the explanation.
    However, no macroscopic U$_7$Te$_{12}$ crystals could be found in the batch, and the phase did not appear in powder x-ray patterns. On the other hand, magnetization measurements sometimes revealed traces of U-Te binary compounds on the tellurium-rich side of UTe$_2$, including U$_7$Te$_{12}$. This is yet another similarity that these crystals possess to those grown in a NaCl/KCl flux \cite{Sakai2022}.
    These observations suggests that U$_7$Te$_{12}$ may be forming as expected but then transforming into UTe$_2$ at lower temperatures, creating crystals in a uranium-rich environment that is thought to be favorable to superconductivity in this material. In any case, the existence of a second synthesis route to high RRR samples of UTe$_2$ should provide valuable opportunities to study the microscopic factors responsible for the differences in T$_c$ in this system.
    
    These Te-flux grown, superconducting crystals where characterized by x-ray diffraction, which confirmed that the crystal structure was that of UTe$_2$, with resistivity measurements, which showed superconductivity at 2K and substantially enhanced RRRs, and with heat capacity measurements, which confirmed the bulk nature of the superconducting transition (see Fig \ref{fig:R_and_C}). 

    %Describe methods
    Resistivity data were collected with a conventional four-terminal, AC lock-in technique. Heat capacity was measured by a quasiadiabatic technique where the calorimeter and sample where heated by about one percent of the bath temperature and allowed to relax. The heat capacity was then extracted by fitting the the rising and falling curves to an exponential function.Thermal conductivity measurements were performed by a one heater, two thermometer, DC method. The typical thermal gradient was between one and two percent of the bath temperature. Contacts to the samples where made with a bismuth-tin-copper solder that maintains good thermal conductivity throughout the temperature and field range studied.

    %describe samples

    % describe fitting
    To study the $T \rightarrow 0$ behavior of the thermal conductivity it is necessary to extrapolate our $\kappa(T)/T$ curves to zero kelvin. The fits shown in Fig \ref{fig:Kappa_v_H} panels a-c were done by finding the best fit to the data of a single power law, $\kappa/t \sim T^{n}$ where both $n$ and the intercept were allowed to vary. For some curves, a good fit to a single power law could only be obtained over a limited temperature range. This was especially true for the highest RRR sample at low fields. Because of this, the fittings were generally done for data below 300mK, and were restricted to even lower temperatures if no visually acceptable fit could be found. Finally, although the zero temperature extrapolations are useful for modeling and conveying the data, we would like to emphasize that all of the trends in the data that support the conclusion of point nodes are visually obvious in the raw data and so the conclusion that point nodes are present in the gap does not depend on the details of these extrapolations.

\begin{figure}
\includegraphics[width=8.5cm]{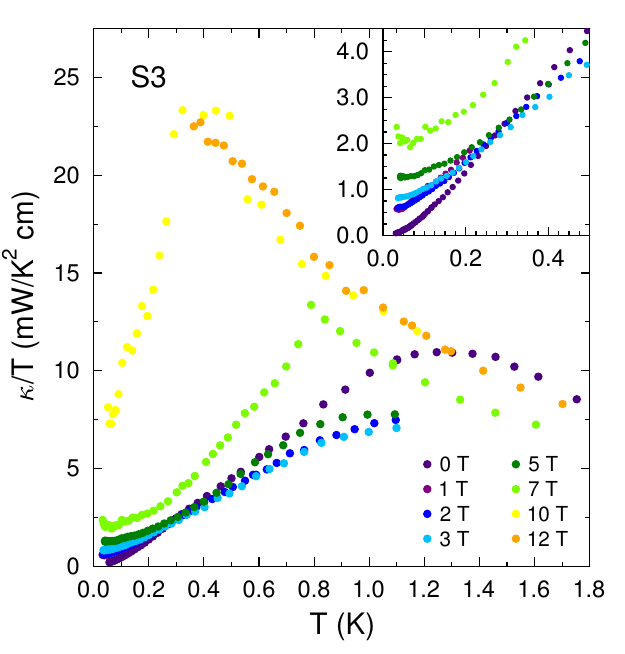} 
\caption{{\bf High field $a-$axis thermal conductivity of sample S3.} Compared to the lower RRR samples, S3 shows dramatically higher thermal conductivity in the normal state at low temperatures, consistent with its higher extrapolated electrical conductivity. Magnetic field is still aligned with the crystal $a-$axis.}
\label{fig:R_and_C} 
\end{figure}
    
\section{Theoretical details.}
% \begin{itemize}
% \item give expression for thermal conductivity in unitary p-wave states used to evaluate theoretical results
% \item refer reader to references for disorder self-energy
% \item give estimate for power law at low $T$ for inelastic scattering with point nodes, and give exact approximation used for inel relaxation rate.
% \item give expression for phonon lifetime used, discuss reason for relying on point scatterers, incl. estimate of  scattering rate from boundary scattering.
% \item Give bare details of field-dependence of $\kappa(T=0,H)$ and parameters used.
% \end{itemize}
\subsection*{Thermal conductivity for a superconductor with point nodes}
The simple model order parameter for a unitary superconductor with point nodes  is given by  an axial state.  On a spherical Fermi surface, the axial state is given by
\begin{equation}
\Delta_\theta = \Delta_0 \sin\theta e^{i\phi},
\end{equation}
where $\theta$ is the polar angle, $\phi$ is the azimuthal angle and $\Delta_0$ is the superconducting gap maximum. This state has point nodes along the $\hat{z}$ direction at the north and south poles of the Fermi surface.  The diagonal thermal conductivity for the axial state \cite{Hirschfeld1988} is given by
%\begin{widetext}
\begin{eqnarray}
\frac{\kappa_{ii}}{T} &=& \hbar k_B^2  \int_{-\infty}^{\infty} d\omega \frac{\omega^2}{T^2} \left(- \frac{d n_{F} (\omega) }{d\omega}\right)  \nonumber \\ 
 &\times & \left\langle  N_0 v_{Fi}^2 \frac{1}{2 \mathrm{Re} \sqrt{\Delta_\theta^2  -\tilde{\omega}^2} }   \left[ 1+ \frac{|\tilde{\omega} |^2 - |\Delta_\theta|^2 }{\left| \Delta_\theta^2  -\tilde{\omega}^2\right|} \right] \right\rangle_{FS},
\label{Eq:kappa_fu}
\end{eqnarray}
%\end{widetext}
where $N_0$ is the single-particle density of states per spin at the Fermi level in the normal state, $v_{Fi}$ is the Fermi velocity along the $i^{th}$ direction, $T$ is the temperature, and $n_F(\omega)$ is the Fermi function. $\langle \dots \rangle_{FS}$ denotes an average over the Fermi surface. The effect of elastic impurity scattering is embedded in $\tilde{\omega}$, which is obtained by solving the following self-consistent equation,
\begin{equation}
\tilde{\omega}=\omega + i 0^+ + \frac{n_{imp}}{\pi N_0} \frac{g_0}{\cot^2 \theta_s - g_0^2}. 
\label{eq:tildew}
\end{equation}
Here $n_{imp}$ is the impurity concentration, $\theta_s\equiv \tan^{-1}(\pi N_0 V_{imp})$ is the $s$-wave scattering phase shift which is related  to the strength of the impurity potential, and $g_0$ is
\begin{equation}
g_0(\tilde{\omega}) = \left\langle \frac{\tilde{\omega}}{\sqrt{\Delta_\theta^2 - \tilde{\omega}^2}} \right\rangle_{FS}.
\label{eq:g0}
\end{equation}
In the zero temperature limit,
\begin{eqnarray}
\frac{\kappa_{ii}}{T}\Big|_{T \rightarrow 0} &=&  \frac{\pi^2 \hbar k_B^2}{3}  \left\langle  N_0 v_{Fi}^2   \frac{1}{2 \mathrm{Re} \sqrt{\Delta_\theta^2  -\tilde{\omega}^2_0} } \right. \nonumber \\
 && \left.  \left[ 1+ \frac{|\tilde{\omega}_0 |^2 - |\Delta_\theta|^2 }{\left| \Delta_\theta^2  -\tilde{\omega}_0^2\right|} \right] \right\rangle_{FS},
%\left\langle  N_0 v_{Fi}^2  \left[  \frac{\tilde{\omega}_0^2 }{(\Delta_0^2 |\dv |^2 +\Gamma_0^2)^{3/2}} \right] \right\rangle_{FS}.
\label{Eq:kappa_fu_T0}
\end{eqnarray}
Here $\tilde{\omega}_0$ is $\tilde{\omega}(\omega=0)$.  In case of point nodes, $\Delta_\theta$ is zero on finite number of points on the Fermi surface, hence, in the absence of any scattering ($\tilde{\omega}_0\rightarrow 0$), $\kappa/T$ vanishes.  { Unlike states with line nodes, for weak disorder an axial state has no residual density of states, nor a limiting low-$T$ term in the thermal conductivity\cite{PJHirschfeld:1986}. Such terms arise only above a critical disorder level depending on the phase shift.} 
{ For the purpose of fitting the thermal conductivity data, we have assumed that the highest $T_{c}$ sample corresponds to $2\%$ reduction in $T_c$, and the impurity concentration is chosen to give this suppression with $\theta_s=\tan^{-1}(3)$. For $T_c=1.8$K and $T_c=1.65K$ samples, the impurity concentrations are chosen to give $15\%$ and $25\%$ $T_c$ suppression, respectively. The  $T_{c}$ is determined by solving,
\begin{eqnarray}
\ln \left( \frac{T_c}{T_{c0}}\right) &=& \Psi\left[ \frac{1}{2}\right] - \Psi\left[ \frac{1}{2} + \frac{\Gamma_N}{2\pi T_{c}}\right].
\end{eqnarray}
Here $\Psi$ denotes the digamma function and the normal state scattering rate $\Gamma_N$ is $({n_{imp}}/{\pi N_0})/{(\cot^2 \theta_s +1 )}$.}
\subsubsection{Inelastic scattering}
As discussed in the main text, the effect of inelastic scattering and contribution from phonons  need to be included in order to describe the experimental data. 
In general, the treatment of simultaneous elastic and inelastic scattering is a complex problem, however here we simply assume that source of scattering is independent, and combine the individual contributions when evaluating thermal conductivity. This can be done by adding the inelastic scattering rate to $\tilde{\omega}$ in Eq. \eqref{Eq:kappa_fu},
\begin{eqnarray}
\tilde{\omega} &\rightarrow & \tilde{\omega} + \frac{i}{2\tau_{inel}}= \mathrm{Re}\left[\tilde{\omega}\right] +\frac{i}{2\tau_{el}} + \frac{i}{2\tau_{inel}}
\end{eqnarray}
The imaginary part of $\tilde{\omega}$ is the elastic scattering rate. We further assume that the origin of the inelastic scattering is a boson that could be phonons or spin-fluctuations, and the bosonic dispersion and the electron-fermion coupling do not have any significant momentum and spin dependence. The low energy bosonic density of states is,
\begin{equation}
\mathcal{D}(\Omega)=\frac{\Omega}{\Omega^2+\Omega_0^2},
\end{equation}
where $\Omega$ is the energy of the boson and $\Omega_0$ is the characteristic energy scale associated with the bosonic mode and it is assumed to be largest energy scale compare to $\Delta_0$, T or $\hbar/2\tau_{el}$. Under these assumptions, the inelastic scattering rate is,
\begin{eqnarray}
\frac{1}{2\tau_{inel}}&=&  \frac{g_{bf}^2}{2} \int_{-\infty}^{\infty}d\Omega N(\omega-\Omega) \mathcal{D}(\Omega)\nonumber \\
& & \times \left[ \coth \left(\frac{\Omega}{2T}\right) + \coth \left(\frac{\omega-\Omega}{2T}\right) \right],
\end{eqnarray}
where $g_{bf}$ is the effective electron-boson coupling constant and $N(\omega)$ is the quasi-particle density of states. In the zero temperature limit, the inelastic scattering rate reduces to
\begin{equation}
\frac{\hbar}{2\tau_{inel}}=  {g_{bf}^2} \int_{0}^{\omega}d\Omega  N(\omega-\Omega) \mathcal{D}(\Omega).
\end{equation}
At low energies $\omega\leq \Delta_0$, the density of states can be approximated as power law $N(\omega)\sim N_0 (\omega/\Delta_0)^\alpha $, and the bosonic density of states can be approximated as $\Omega/\Omega_0^2$ and the inelastic lifetime becomes,
\begin{equation}
\frac{\hbar}{2\tau_{inel}} \approx \frac{N_0 g_{bf}^2}{\Omega_0^2 \Delta_0^\alpha (\alpha^2 + 3\alpha+2)} \omega^{\alpha+2}. 
\end{equation}
For line nodes, $\alpha=1$ leads to $\frac{1}{2\tau_{inel}} \propto \omega^3$ and for point nodes $\alpha=2$ that gives $\frac{1}{2\tau_{inel}} \propto \omega^4$. The integrand in Eq. \eqref{Eq:kappa_fu} is sharply peaked at $\omega=0$, and observing the weak $\omega$ dependence of inelastic scattering rate, we only retain the temperature dependence on $\tau_{inel}^{-1}$ at $\omega=0$. At $\omega=0$, the inelastic scattering rate becomes,
\begin{equation}
\frac{\hbar}{2\tau_{inel}}=  {g_{bf}^2} \int_{0}^{\infty}d\Omega \frac{ N(-\Omega) \mathcal{D}(\Omega)}{\sinh(\Omega/T)}.
\end{equation}
This integral is also peaked at $\Omega=0$, therefore, it can be approximated as,
\begin{equation}
\frac{\hbar}{2\tau_{inel}} \approx  \frac{{g_{bf}^2}}{\Omega_0^2} \int_{0}^{\infty}d\Omega \frac{ N(\Omega) \Omega}{\sinh(\Omega/T)},
\end{equation}
which yields $T^4$ temperature dependence for point nodes and $T^3$ for line nodes. The prefactor ${{g_{bf}^2}}/{\Omega_0^2} $ can be fixed by the value of the inelastic scattering rate at $T_c$ and it is used as a fitting parameter. We find that the inelastic scattering  rate ${\hbar}/{2\tau_{inel}}=0.2T_{c0}$ at $T_{c0}$ gives a reasonable fit to the experimental data.   Apart from the inelastic scattering rate, the electron-boson scattering also affects the effective mass. However, the correction to the effective mass is weak, noting that the effective mass is quite high in UTe$_2$.
\subsubsection{Phonon thermal conductivity}
The general expression for the phonon thermal conductivity is\cite{Uher1990,BRT1959},
\begin{eqnarray}
\kappa_{ph} &=& \frac{k_B^4 T^3 }{2\pi^2 \hbar^3 v_s^4}\int_{0}^{\infty}dx \frac{x^4 e^x}{(e^x-1)^2} \ell(x,T), 
\end{eqnarray}
where $v_s$ is the sound velocity,  $x$ is $\hbar \omega/k_B T$, $\omega$ is the phonon energy and $\ell(x,T)$ is the effective mean free path given by,
\begin{eqnarray}
\frac{1}{\ell(x,T)}&= & \frac{1}{\ell_B}+ \frac{1}{\ell_{d}}+ \frac{1}{\ell_{e-ph}} + \frac{1}{\ell_U},
\end{eqnarray}
where $\ell_{B}$, $\ell_d$, $\ell_{e-ph}$ and $\ell_{U}$ are the mean free path due to boundary, defect, electron-phonon, Umklapp scattering processes, respectively. 

In order to determine the relevant scattering mechanism for phonons below T$\lesssim$2K, we need to estimate the sound velocity and Debye temperature. Within the Debye model the low temperature specific heat due to phonons is given by $C_{ph} (T)=A_{ph} T^3$, where $A_{ph}$ is,
\begin{equation}
A_{ph}=\frac{12\pi^4}{5} N_{atoms} k_B \frac{1}{\Theta_D^3}.
\end{equation}
Here $N_{atoms}$ is the number of atoms per mole, and $\Theta_D$ is the Debye temperature. Using the experimentally estimated value $A_{ph}=2.84$mJK$^{-4}$mole$^{-1}$\cite{metz_point-node_2019} and $N_{atoms}=$1.8$\times$10$^{24}$/mole, we find $\Theta_D=$127K. The sound velocity $v_s\equiv \omega_D \sqrt[3]{V_{u.c.}/6\pi^2 N_{p.u.c}}$ is 1.3km/s using the unit cell volume $V_{u.c.}=$355$\mathrm{\AA}^3$, and number of atoms per unit cell $N_{p.u.c}=$12\cite{Aoki_2022_Review}. We are interested in phonon thermal conductivity below T$_c\ll \Theta_D$, therefore, Umklapp scattering can be ignored. There is no evidence for strong electron-phonon coupling\cite{YangLee2023}, and in the superconducting state $\ell_{e-ph}$ is expected to be very large due to reduced quasi-particle density of states\cite{Tewordt1989}.  Hence, the boundary scattering $\ell_B$ and scattering by the point-defects $\ell_d$ are the two main mechanisms that determine the phonon mean free path. The boundary scattering is determined by the sample size and the amount of diffusive scattering by the boundaries. On the other hand, the scattering by the point defects is an energy dependent process. The mean free path due to point defects is given by\cite{Klemens1955},
\begin{equation}
\frac{1}{\ell_d(x,T) }=  \left[ \frac{81\pi^4}{4 V_{u.c.} } N_{p.u.c.}^4 \right]^{\frac{1}{3}} \left(\frac{T}{\Theta_D}\right)^4 x^4 \sum_{i} N_{di} \left( \frac{\delta M_i}{M_i}\right)^2. 
\label{Eq:ld}
\end{equation}
Here $N_{di}$ is the number of $i^{th}$ kind of defect, $M_i$ is the mass of $i^{th}$ kind of ion and $\delta M_i$ is the mass difference of kind of defect from $M_i$. For a vacancy $\delta M_i= M_i$ and for a different type of isotope $\delta M_i= M_i-M_{isotope}$, where $M_{isotope}$ is the isotope mass. We rewrite the $\kappa_{ph}$,
\begin{equation}
\frac{\kappa_{ph}}{T} = \alpha_0 T^2 \int_{0}^{\infty}dx \frac{x^4 e^x}{(e^x-1)^2} \frac{1}{1+\alpha_1 T^4 x^4 },
\end{equation}
where $\alpha_0$ and $\alpha_1$ are,
\begin{eqnarray}
\alpha_0 &=& \frac{k_B^4  \ell_B  }{2\pi^2 \hbar^3 v_s^4}, \\
\alpha_1 &=&  \left[ \frac{81\pi^4}{4 V_{u.c.} } N_{p.u.c.}^4 \right]^{\frac{1}{3}} \frac{\ell_B }{\Theta_D^4}  \sum_{i} N_{di} \left( \frac{\delta M_i}{M_i}\right)^2.
\end{eqnarray}
%\begin{eqnarray}
%\kappa_{ph} &=& \frac{k_B^4 T^3 \ell_B  }{2\pi^2 \hbar^3 v_s^4}\int_{0}^{\infty}dx \frac{x^4 e^x}{(e^x-1)^2} \frac{1}{1+\frac{\ell_B}{\ell_d(x,T)}}.
%\label{eq:kph2}
%\end{eqnarray}
Since it is not possible to determine the effective boundary scattering length in the presence of realistic boundaries causing diffusive scattering, or the number of defects,  we use $\alpha_0$ and $\alpha_1$ as independent two fitting parameters to capture the effect of these two types of phonon mean free paths. Table \ref{tab:ph_fit_param} shows the parameters $\alpha_0$ and $\alpha_1$ used to fit the data.

\begin{table}[tbph]
    \centering
    \begin{tabular}{cccc}
    \hline 
    $T_c$ (K) & $\alpha_0$ (mW/K$^{4}$/cm) & $\alpha_1$ (K$^{-4}$) & $ \kappa_{ph}/\kappa_{el}|_{T=T_c}$ \\
    \hline
     2.1 & 1.92 & 0.113 & 0.89 \\
     1.8 & 1.57 & 0.124 & 1.16 \\
     1.65& 0.84 & 0.151 & 0.74 \\
     \hline
    \end{tabular}
    \caption{Parameters used to calculate the phonon thermal conductivity for each sample.}
    \label{tab:ph_fit_param}
\end{table}

\subsubsection{Magnetic field dependent thermal conductivity}
Within the semi-classical formalism\cite{Volovik1993,Kuebert1998,Matsuda2006}, the effect of the magnetic field can be described by shifting the quasiparticle energy by a Doppler shift, which is,
\begin{equation}
\delta \omega = \frac{1}{2} m^\ast \mathbf{v}_F\cdot \mathbf{v}_s,
\end{equation}
where $\mathbf{v}_s$ is the superfluid velocity, $\mathbf{v}_F$ is the Fermi velocity, and $m^\ast$ is the effective mass. The superfluid velocity is,
\begin{eqnarray}
\mathbf{v}_s = \frac{\hbar}{2 m^\ast r}\hat{\Psi},
\end{eqnarray}
where $r$ is the distance from the vortex core in the real space and $\Psi$ is the winding angle. We rewrite the Doppler shift energy as,
\begin{eqnarray}
\delta \omega &=& \Delta_0 \frac{1}{\rho} \mathcal{B} \sqrt{\frac{\pi \eta}{16}} \sqrt{\frac{H}{H_{c2}}} f(\phi,\theta,\Psi),
\label{eq:ds2}
\end{eqnarray}
where $\rho\equiv r/R_H$ is the distance from the vortex core in the units of magnetic length scale $R_H\equiv \sqrt{\Phi_0/\eta H}$. Here $\Phi_0$ is the magnetic flux quanta, and $\eta$ is a dimensionless parameter of the order of unity. The value of $\eta$ depends on the  vortex lattice  structure and we use $\eta=\sqrt{3}/4 $ which correspond to a triangular vortex lattice, another dimensionless  parameter $\mathcal{B}=\hbar |v_F|/\pi \Delta_0 \xi_0$ is used as a free parameter in our calculations and we have set $\mathcal{B}=2.5$. The upper critical field $H_{c2}\equiv \Phi_0/\pi \xi_0^2$. We have assumed that the upper critical fields are isotropic along all directions and the Fermi surface is a sphere. 
% To study the  low field qualitative behavior, our assumption is reasonable.
{ The true anisotropy of the system can in principle be accounted for by assuming  anisotropic coherence lengths, Fermi velocities, and including the effects of disorder on these quantities, which we have not done, in part because we were analyzing data for $\bf J_{Q},H \parallel \hat a$ only.   }
In Eq. \eqref{eq:ds2}, function $f(\phi,\theta,\Psi)$ contains the information about the magnetic field direction and the Fermi surface topology. For a magnetic field along the nodes $\hat{z}$ axis and perpendicular to the nodes $\hat{x}$, $f$ are,
\begin{eqnarray}
f(\phi,\theta,\Psi)_{H \parallel \hat{z} } &=& \sin \theta  \sin \left( \phi - \Psi \right) \\
f(\phi,\theta,\Psi)_{H \parallel \hat{x} } & =&  \left[ \cos \Psi \sin \theta \sin \phi + \sin \Psi \cos \theta\right].
\end{eqnarray}
In order to calculate the effect magnetic field on thermal conductivity, we first solve Eq. \eqref{eq:tildew} with Doppler shifted quasiparticle energy $\omega\rightarrow \omega-\delta \omega$. This makes impurity renormalized quasi-particle energy a function of $\rho$, $\Psi$. Now, using $\tilde{\omega}(\rho,\Psi)$, we calculate $\kappa(\rho,\Psi)/T$, and finally, we   average $\kappa$ over the vortex lattice unit cell. The averaging over the vortex lattice unit cell is defined as,
\begin{equation}
\kappa = \frac{1}{\mathbf{A}} \int_{0}^{1}d\rho \rho \int_0^{2\pi}d\Psi \kappa(\rho,\Psi),
\end{equation}
where $\mathbf{A}$ is the vortex lattice unit cell area.
%\begin{figure*}
%\includegraphics[width=5.9cm]{TFit_2p1K_t5_Tcr_2.png} 
%\includegraphics[width=5.9cm]{TFit_1p8K_t5_Tcr_15.png} 
%\includegraphics[width=5.9cm]{TFit_1p6K_t3_Tcr_25.png} 
%\includegraphics[width=5.9cm]{HFit_2p1K_t5_local_Tcr_2.png} 
%\includegraphics[width=5.9cm]{HFit_1p8K_t5_local_Tcr_15.png} 
%\includegraphics[width=5.9cm]{HFit_1p6K_t5_local_Tcr_25.png} 
%\caption{{\bf Field dependence of the normalized thermal conductivity for an axial %state on a spherical Fermi surface. }.}
%\label{fig:Theory3A} 
%\end{figure*}

\bibliography{references}

\end{document}